\documentclass[aps,pre,reprint,showpacs,superscriptaddress,longbibliography]{revtex4-1}
\usepackage{amsmath,amssymb,graphicx,units}
\usepackage[usenames,dvipsnames]{color}

\begin{document}
\title{Eigenstate thermalization in the two-dimensional transverse field Ising model}
\author{Rubem Mondaini}
\affiliation{Department of Physics, The Pennsylvania State University, University Park, Pennsylvania 16802, USA}
\affiliation{Beijing Computational Science Research Center, Beijing 100193, China}
\author{Keith R. Fratus}
\affiliation{Department of Physics, University of California, Santa Barbara, California, 93106, USA}
\author{Mark Srednicki}
\affiliation{Department of Physics, University of California, Santa Barbara, California, 93106, USA}
\author{Marcos Rigol}
\affiliation{Department of Physics, The Pennsylvania State University, University Park, Pennsylvania 16802, USA}

\begin{abstract}
We study the onset of eigenstate thermalization in the two-dimensional transverse field Ising model (2D-TFIM) in the square lattice. We consider two non-equivalent Hamiltonians: the ferromagnetic 2D-TFIM and the antiferromagnetic 2D-TFIM in the presence of a uniform longitudinal field. We use full exact diagonalization to examine the behavior of quantum chaos indicators and of the diagonal matrix elements of operators of interest in the eigenstates of the Hamiltonian. An analysis of finite size effects  reveals that quantum chaos and eigenstate thermalization occur in those systems whenever the fields are nonvanishing and not too large.
\end{abstract}

\maketitle

\section{Introduction}
The transverse field Ising model (TFIM) is one of the simplest models that exhibits both ground-state and finite-temperature (in dimensions higher than one) phase transitions between paramagnetic and ordered phases. The three-dimensional TFIM was used by DeGennes to characterize the ferroelectric phase of KH$_2$PO$_2$~\cite{DeGennes1963}, and the one-dimensional TFIM was recently realized in experiments with ultracold bosons in tilted optical lattices~\cite{Simon2011}. This was possible via a mapping of the site occupation of the bosonic atoms onto pseudo-spins~\cite{Sachdev2002}. The one-dimensional TFIM has been extensively studied theoretically in recent years in the context of quantum quenches in integrable systems \cite{rossini_silva_09, rossini_susuki_10, calabrese_essler_11, foini_cugliandolo_12, calabrese_essler_12a, calabrese_essler_12b}. The two-dimensional TFIM (2D-TFIM), on the other hand, is not integrable. It was examined by two of us (KRF and MS) \cite{FratusSrednicki} to understand whether eigenstate thermalization \cite{deutsch1991quantum, srednicki1994chaos, rigol_dunjko_08} occurs in the presence of long-range order. 

Eigenstate thermalization is a phenomenon that has received much attention recently as it explains why thermalization occurs in generic isolated quantum systems when taken far from equilibrium \cite{mrigolreview2015}. Specifically, the fact that observables after relaxation can be described using traditional ensembles of statistical mechanics has been argued to be the result of the matrix elements of those observables in the eigenstates of the Hamiltonian being equal to the thermal expectation values \cite{deutsch1991quantum, srednicki1994chaos, rigol_dunjko_08}. Another way to state this is that the eigenstate to eigenstate fluctuations of the expectation values of the observables is very small, more precisely, exponentially small in the system size \cite{mrigolreview2015}. Many studies of quantum systems, mainly in one-dimensional lattices, have found results consistent with this \cite{rigol_09a, rigol_09b, santos_rigol_10b, neuenhahn_marquardt_12, genway_ho_12, khatami_pupillo_13, beugeling_moessner_14, kim_14, sorg_vidmar_14}. Eigenstate thermalization can be understood as being a result of quantum chaos \cite{mrigolreview2015}, and indeed the onset of eigenstate thermalization has been seen to coincide with the onset of quantum chaos in some one-dimensional systems \cite{santos_rigol_10b,santos_rigol_10a}.

In this work, we present an in depth study of quantum chaos and eigenstate thermalization indicators in the 2D-TFIM in the square lattice. In contrast to the study in Ref.~\cite{FratusSrednicki}, we do not introduce any symmetry breaking perturbation in the Hamiltonian to discern order. Instead, we use structure factors, which reveal order even in the absence of symmetry breaking. Also, in addition to the ferromagnetic 2D-TFIM considered in Ref.~\cite{FratusSrednicki}, here we study the antiferromagnetic 2D-TFIM in the presence of a longitudinal field. We study both models in various clusters with periodic boundary conditions, which allows us to present a finite size scaling analysis of the quantities of interest.

The presentation is organized as follows: In Sec.~\ref{sec:sec2}, we introduce the model and discuss the numerical approach used to study it. Section~\ref{sec:sec3} is devoted to the analysis of quantum chaos indicators and their scaling. Section~\ref{sec:sec4} is devoted to the analysis of eigenstate thermalization indicators and their scaling. A summary of the results are presented in Sec.~\ref{sec:sec5}. 

\section{Model and Numerical approach}\label{sec:sec2}

The Hamiltonian of the 2D-TFIM in the presence of a longitudinal field can be written as,
\begin{equation}
 \hat H = J\sum_{\langle {\bf i},{\bf j}\rangle}\hat\sigma_{\bf i}^z\hat\sigma_{\bf j}^z + g\sum_{\bf i}\hat\sigma_{\bf i}^x + \varepsilon \sum_{\bf i} \hat\sigma_{\bf i}^z,
 \label{eq:hamiltonian}
\end{equation}
where $\hat\sigma_{\bf i}^z$ and $\hat\sigma_{\bf i}^x$ are the $z$ and $x$ Pauli matrices, respectively, at site ${\bf i}$ of the lattice. $J$ is the strength of the nearest neighbor ($\langle {\bf i},{\bf j}\rangle$ in the summation) Ising interaction. We consider both the ferromagnetic ($J<0$) and the antiferromagnetic ($J>0$) cases, and set $|J|=1$ as our energy scale. $g$ and $\varepsilon$ are the strength of the transverse and longitudinal fields, respectively. We denote the total number of sites in the system by $N$. 

First, it is important to mention some symmetries of this model in the square lattice, which is a bipartite lattice. In the absence of the longitudinal field ($\varepsilon=0$), the ferromagnetic and the antiferromagnetic 2D-TFIMs are connected through the transformation $\hat\sigma_{\bf i}^z\rightarrow (-1)^{i_x+i_y}\hat\sigma_{\bf i}^z$. This transformation maps the uniform magnetization per site $M = \langle \sum_{\bf i} \sigma_{\bf i}^z \rangle/N$, which is the order parameter in the ferromagnetic case, onto the staggered magnetization per site $M_\text{stag} = \langle \sum_{\bf i}(-1)^{i_x+i_y} \sigma_{\bf i}^z\rangle/N$, which is the order parameter in the antiferromagnetic case, and vice versa. Thus, the phase transitions in both models occur at the same values of $g$. For this reason, for $\varepsilon=0$, in this work we study only the ferromagnetic case. (The ground-state phase transition separating the paramagnetic and ordered phases occurs at a critical transverse field $g_c\simeq3.044$~\cite{Rieger1999}.) We note that this model has a $Z_2$ symmetry associated with its invariance under the transformation $\hat\sigma_{\bf i}^z\rightarrow -\hat\sigma_{\bf i}^z$.  In addition, here we study the antiferromagnetic 2D-TFIM in the presence of a uniform longitudinal field. We restrict our analysis to the case $\varepsilon=g$. This model is equivalent to the ferromagnetic 2D-TFIM in the presence of a staggered longitudinal field, which breaks the $Z_2$ symmetry of the model with $\varepsilon=0$.

\begin{figure}[!tb] 
  \includegraphics[width=0.95\columnwidth]{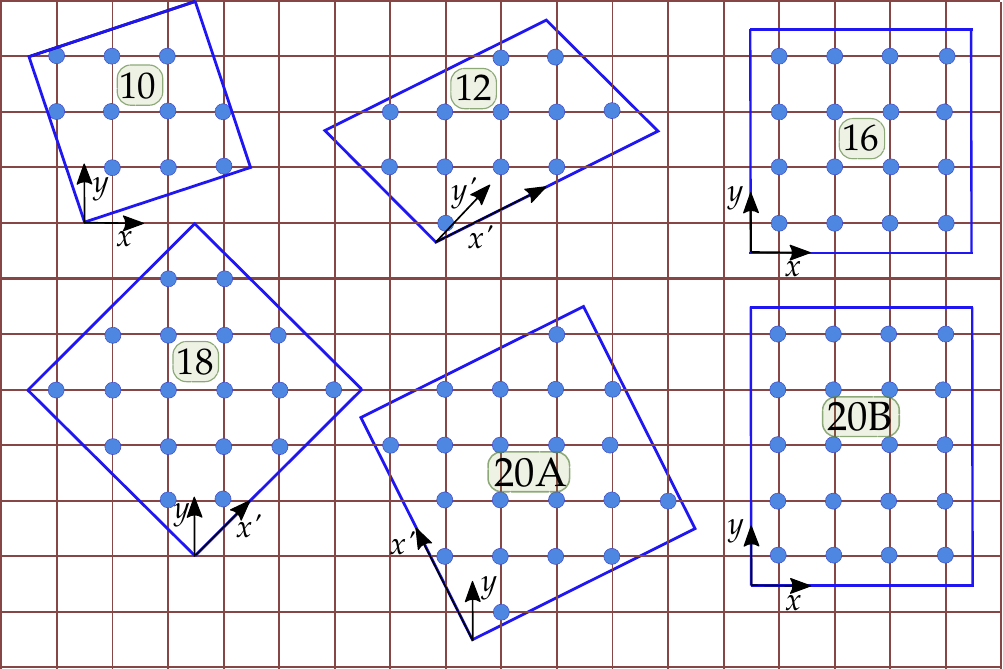}
 \vspace{-0.1cm}
 \caption{(Color online) Clusters with periodic boundary conditions used in this work. All clusters, with the exception of the non-tilted lattice with 20 sites (bottom right), support the N\'eel state. Each cluster displays the basis in which translation symmetry operations are implemented.}
 \label{fig:tilted_square}
\end{figure}

In order to study quantum chaos indicators and calculate the expectation values of observables in eigenstates of the Hamiltonian, we use full exact diagonalization of clusters with different sizes and periodic boundary conditions. All the clusters considered in this work are shown in Fig.~\ref{fig:tilted_square}. Most of them have a tilted structure that is needed to accommodate the N\'eel state \cite{dagotto_review_94}, which is the ground state of the antiferromagnetic Ising model. 

\begin{table}
\caption{Dimension ${\mathcal D}$ of the Hilbert subspaces for the different clusters in Fig.~\ref{fig:tilted_square} after the breakup in the $Z_2$ and momentum sectors. In the left columns, the number in the first (second) parenthesis is the size of the odd (even) subspace associated with the $Z_2$ symmetry. Momentum sectors are indicated in the right column. There are momentum sectors that exhibit spatial symmetries. We have only implemented inversion. Whenever there are two numbers inside parentheses in the first column, the first (second) number indicates the size of the odd (even) subspace associated with the inversion symmetry. The axes used for the translations (in $n_x$, $n_y$, $n_{x'}$, and $n_{y'}$) are indicated in Fig.~\ref{fig:tilted_square}.}
\begin{center}
\begin{tabular}{l r}
\hline \hline 
$N=10$&$(k_x,k_y)=\frac{\pi}{5}(n_x,n_y)$ \\
\hline \hline 
${\mathcal D}$&$(n_x,n_y)$\\
\hline
(18+34)+(12+44)&$(0,0)$ \\
(34+18)+(24+24)&$(5,0)$ \\
(51)+(48)&$(1,0);(3,0);(7,0);(9,0)$ \\
(51)+(54)&$(2,0);(4,0);(6,0);(8,0)$ \\
\hline\hline 
$N=12$&$(k_x,k_y)=$\\ 
      & $\frac{\pi}{3}(3n_{x^\prime}-n_{y^\prime},-3n_{x^\prime}+2n_{y^\prime})$ \\
\hline \hline 
${\mathcal D}$&  $(n_{x^\prime},n_{y^\prime})$\\
\hline
(70+102)+(55+135)& $(0,0)$\\
(70+102)+(75+91)& $(0,3);(1,0);(1,3)$\\
(170)+(165)& $(0,1);(0,5);(1,1)$\\
           & $(1,2);(1,4);(1,5)$\\
(170)+(185)& $(0,2);(0,4)$\\
\hline\hline 
$N=16$  &  $(k_x,k_y)=\frac{\pi}{2}(n_x,n_y)$ \\
\hline \hline 
${\mathcal D}$&  $(n_x,n_y)$\\
\hline
(960+1088)+(894+1214) & $(0,0)$\\
(960+1088)+(1014+1078) & $(2,2)$\\
(1088+960)+(1078+1014) & $(0,2); (2,0)$\\
(2048)+(2032) & $(0,1); (1,0); (0,2);(2,0)$\\ 
              & $(0,3); (3,0); (1,2); (2,1)$\\ 
              & $(1,3); (3,1); (2,3); (3,2)$\\
\hline\hline 
$N=18$  &   $(k_x,k_y)=\frac{\pi}{3}(2n_{x^\prime}-n_y,n_y)$ \\
\hline \hline 
${\mathcal D}$&  $(n_{x^\prime},n_y)$\\
\hline
(3520+3776)+(3408+3920)& $(0,0)$\\
(3776+3520)+(3632+3632)& $(0,-3)$\\
(7280)+(7252)& $(0,\pm1);(\pm1,-3);(\pm1,\pm1)$\\
(7280)+(7308)& $(\pm1,0);(\pm1,\pm2);(0,\pm2)$\\
\hline\hline 
$N=20$  &   $(k_x,k_y)=\frac{\pi}{5}(-5n_{x^\prime}+2n_y,n_y)$ \\
\hline \hline 
${\mathcal D}$ &  $(n_{x^\prime},n_y)$\\
\hline
(12852+13364)+(12546+13826) & $(0,0)$\\
(12852+13364)+(12954+13210) & $(1,5)$\\
(13364+12852)+(13210+12954)      & $(0,5);(1,0)$\\
(26214)+(26163)      & $(0,1);(0,3);(0,7);(0,9)$\\
                     & $(1,1);(1,3);(1,7);(1,9)$\\
                     & $(1,2);(1,4);(1,6);(1,8)$\\
(26214)+(26367)           & $(0,2);(0,4);(0,6);(0,8)$\\
\hline\hline 
\end{tabular}
\end{center}
\label{table:dimensions}
\end{table}

We make use of translation symmetry to break up the Hamiltonian in momentum sectors. In addition, for the ferromagnetic model, we breakup each momentum sector using the $Z_2$ symmetry. There are some momentum sectors that exhibit space symmetries. We do not use them all. We only implemented inversion (whenever present). In Table~\ref{table:dimensions}, we show the breakup of the Hilbert space for all the clusters studied. We note that, for the calculations of the antiferromagnetic case, the $Z_2$ symmetry is absent so the linear dimension of all matrices diagonalized was around twice as large as those involved in the calculations of the ferromagnetic case. We also note that the cluster 20B (see Fig.~\ref{fig:tilted_square}) does not accommodate the N\'eel state; besides, it displays larger finite size effects in comparison to cluster 20A (see Appendix). This is why we omit its results in the main text in favor of the ones for lattice 20A.

\section{Quantum chaos indicators}\label{sec:sec3}

\subsection{Distribution of the ratio of consecutive gaps}

We first study the statistics of energy level spacings. A system is said to be quantum chaotic if the distribution of normalized energy level spacings follows a Wigner-Dyson function, which exhibits level repulsion \cite{bohigas_giannoni_84}. On the other hand, as per Berry-Tabor's conjecture \cite{berry1977level}, one expects a Poisson distribution when the system is integrable. To avoid the unfolding procedure of the spectra needed to guarantee that the energy level spacings are normalized to unity, here we use the ratio of the smallest to the largest consecutive energy gaps \cite{oganesyan_huse_07}: $r_n = \min\left(\delta_n,\delta_{n+1}\right)/\max\left(\delta_n,\delta_{n+1}\right)$, where $\delta_n = E_{n+1}-E_n$ and $\{E_n\}$ is the ordered list of eigenenergies in a particular symmetry sector. For quantum chaotic systems with time-reversal symmetry, for which the relevant random matrices belong to the Gaussian orthogonal ensemble (GOE), the distribution of $r$ is given by the expression~\cite{atas2013distribution}:
\begin{equation}\label{eq:pgoe}
 P_\text{GOE}(r) = \frac{27}{4}\frac{r+ r^2}{(1+r+r^2)^{\frac{5}{2}}}\Theta(1-r).
\end{equation}
This distribution is expected to apply to the 2D-TFIM in the quantum chaotic regime as the Hamiltonian \eqref{eq:hamiltonian} can always be written as a real matrix. In integrable regimes, on the other hand, the Poisson distribution results in
\begin{equation}
 P_\text{P}(r)=\frac{2}{1+r^2}\Theta(1-r).
\end{equation}
In quantum chaotic systems, the presence of unresolved symmetries results in a distribution $P(r)$ that is between $P_\text{GOE}(r)$ and $P_\text{P}(r)$.

\begin{figure}[!tb] 
 \includegraphics[width=0.99\columnwidth]{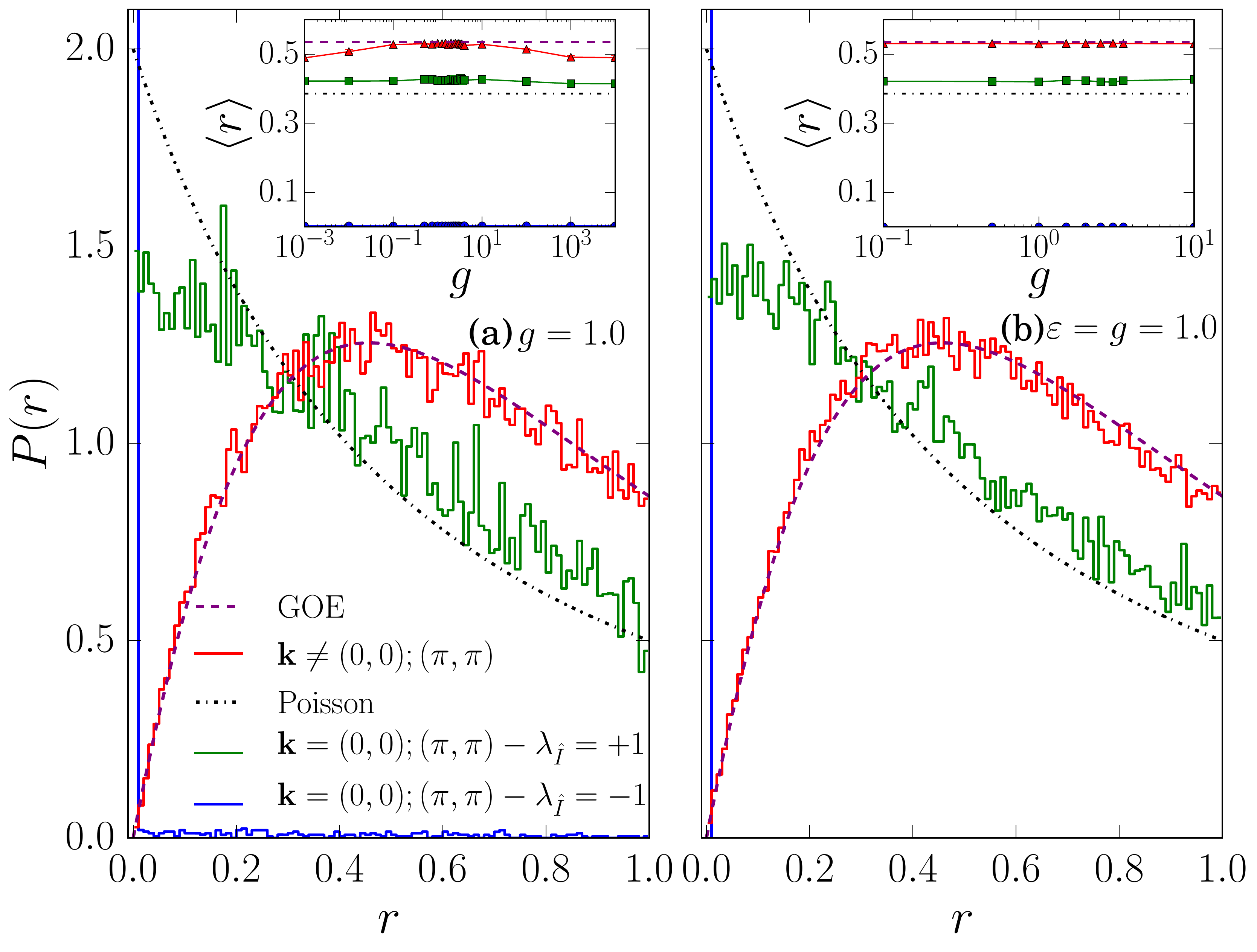}
 \vspace{-0.1cm}
 \caption{(Color online) Distribution of the ratio of consecutive energy gaps in the spectra for: (a) the ferromagnetic case with $g=1.0$ ($\varepsilon = 0$) and (b) the antiferromagnetic case with $\varepsilon=g=1$. The results were obtained in the cluster 20A (see Fig.~\ref{fig:tilted_square}). Results are reported for $P(r)$ averaged over all momentum sectors with $\textbf{k}\neq(0,0)$ and $\textbf{k}\neq(\pi,\pi)$, in which $Z_2$ (for the ferromagnetic case) and parity under inversion [for $\textbf{k}=(0,\pi)$ and $\textbf{k}=(\pi,0)$] are the only additional symmetries and they are resolved. We also show the average $P(r)$ between the momentum sectors $\textbf{k}=(0,0)$ and $\textbf{k}=(\pi,\pi)$ when divided in the even ($\lambda_{\hat I}=+1$) and odd ($\lambda_{\hat I}=+1$) parity sectors under inversion. In those momentum sectors inversion is not the only space symmetry. (Insets) Average value of $r$ as a function of the strength of the fields. The horizontal dashed lines depict the average predicted by $P_\text{GOE}(r)$ (top) and $P_\text{P}(r)$ (bottom). All results were obtained using the the central half of the spectrum in each subspace.}
  \label{fig:P_r_AF_and_F}
\end{figure}

Figure~\ref{fig:P_r_AF_and_F} shows the numerical results obtained for $P(r)$ averaged over all momentum sectors excluding $\textbf{k}=(0,0)$ and $\textbf{k}=(\pi,\pi)$. In the latter two sectors inversion is not the only space symmetry. In Fig.~\ref{fig:P_r_AF_and_F}(a), we report results for the ferromagnetic case and, in Fig.~\ref{fig:P_r_AF_and_F}(b), for the antiferromagnetic case. They are in very good agreement with $P_\text{GOE}(r)$. We should add that $P_\text{GOE}(r)$ in Eq.~\eqref{eq:pgoe} was obtained for $3\times3$ matrices, and is expected to be slightly different in the thermodynamic limit \cite{atas2013distribution}. Our results indicate that, in the thermodynamic limit, $P_\text{GOE}(r)$ is slightly larger (smaller) than in Eq.~\eqref{eq:pgoe} for $r$ smaller (larger) than the value for which $P_\text{GOE}(r)$ is maximal, in agreement with the analysis in Ref.~\cite{atas2013distribution}.

In the momentum sectors with $\textbf{k}=(0,0)$ and $\textbf{k}=(\pi,\pi)$, Fig.~\ref{fig:P_r_AF_and_F} shows that $P(r)$ is in between $P_\text{GOE}(r)$ and $P_\text{P}(r)$ in the even parity sector under inversion ($\lambda_{\hat I}=+1$). On the other hand, in the odd parity sector under inversion ($\lambda_{\hat I}=-1$), we find that there are pairs of degenerate states across the spectrum, which results in a $\delta$-like peak in $P(r)$ at $r\approx0$. This highlights the importance of resolving all symmetries for one to be able to identify the presence of quantum chaos in the distribution of level spacings. We note that, the highly symmetric clusters with 16 and 18 sites [$P(r)$ is not shown for those clusters] exhibit space symmetries (not necessarily inversion) in all momentum sectors.

The insets in Fig.~\ref{fig:P_r_AF_and_F} display the average value of $r$ as a function of the strength of the fields in the sectors with $\textbf{k}\neq(0,0)$ and $\textbf{k}\neq(\pi,\pi)$, in which all symmetries are resolved. We plot as horizontal dashed lines the predictions of $P_\text{GOE}(r)$, $\langle r\rangle_\text{GOE}=0.5359$, and of $P_\text{P}(r)$, $\langle r\rangle_\text{P} =2\ln2-1 \approx0.386$~\cite{atas2013distribution}. Note that, away from the integrable limits $g=0$ and $g=\infty$, the results are consistent with $\langle r\rangle_\text{GOE}$. For $\textbf{k}=(0,0)$ and $\textbf{k}=(\pi,\pi)$, $\langle r\rangle$ is close to $\langle r\rangle_\text{P}$ for all values of $g$ studied. It is worth stressing that, given the fact that our Hamiltonian contains only short-range interactions, the GOE prediction is valid only away from the edges of the spectrum \cite{brody1981random, flores2001spectral, kaplan2000wave, santos_rigol_10a, santos_rigol_10b}. This, and to minimize finite-size effects, is why all results reported in Fig.~\ref{fig:P_r_AF_and_F} were obtained using the central half of the spectrum in each subspace analyzed.

\begin{figure*}[!t] 
 \includegraphics[width=1.98\columnwidth]{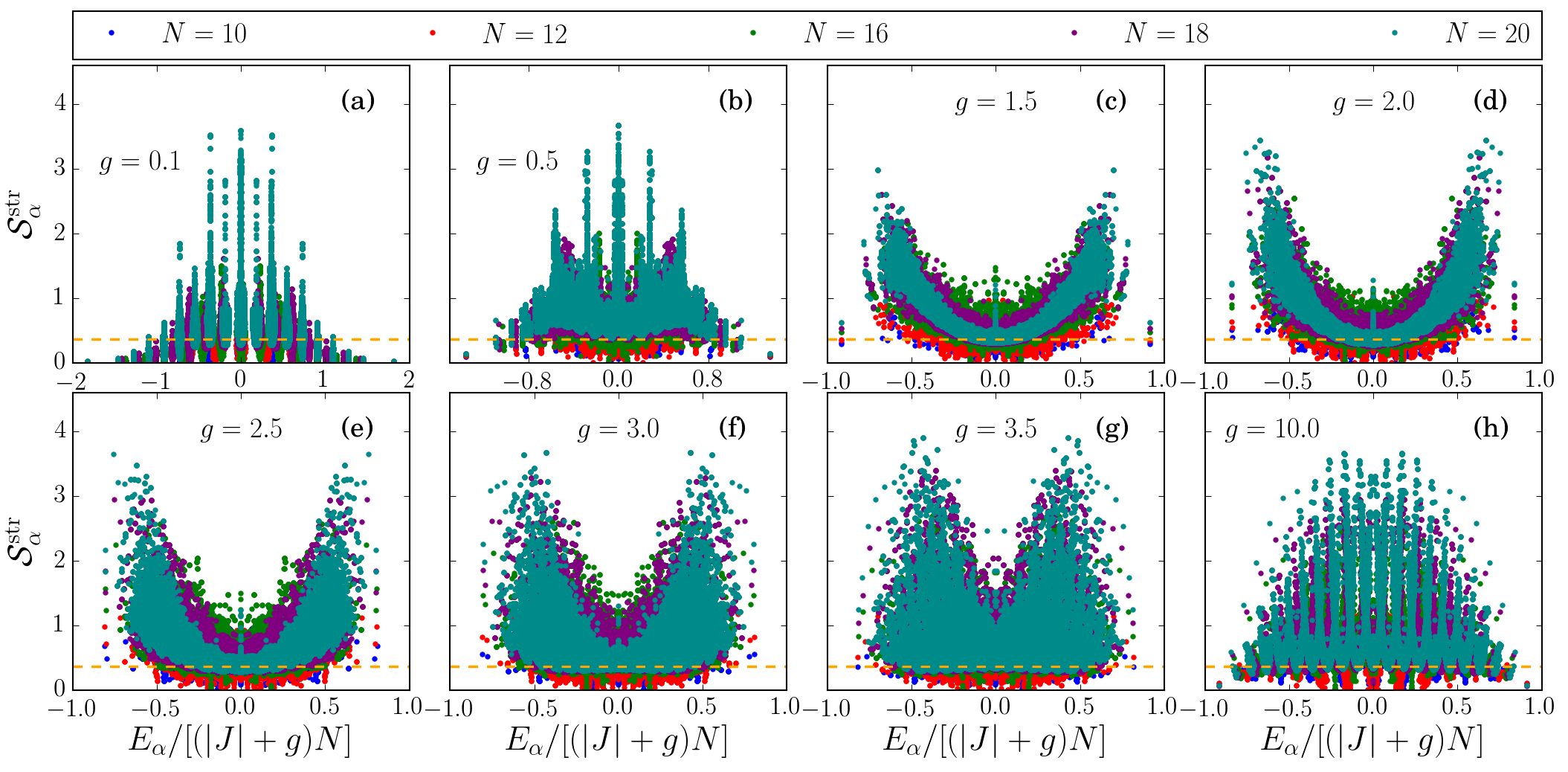}
 \vspace{-0.1cm}
 \caption{(Color online) Structural entropy in all symmetry sectors of the ferromagnetic 2D-TFIM ($J=-1$ and $\varepsilon=0$) for different system sizes (see Table~\ref{table:dimensions}). The narrowing of the support of the values of the structural entropy with increasing system size, in any given energy window, is an indication of the occurrence of quantum chaos.}
  \label{fig:struct_entropy_F}
\end{figure*}
\begin{figure*}[!t] 
 \includegraphics[width=1.98\columnwidth]{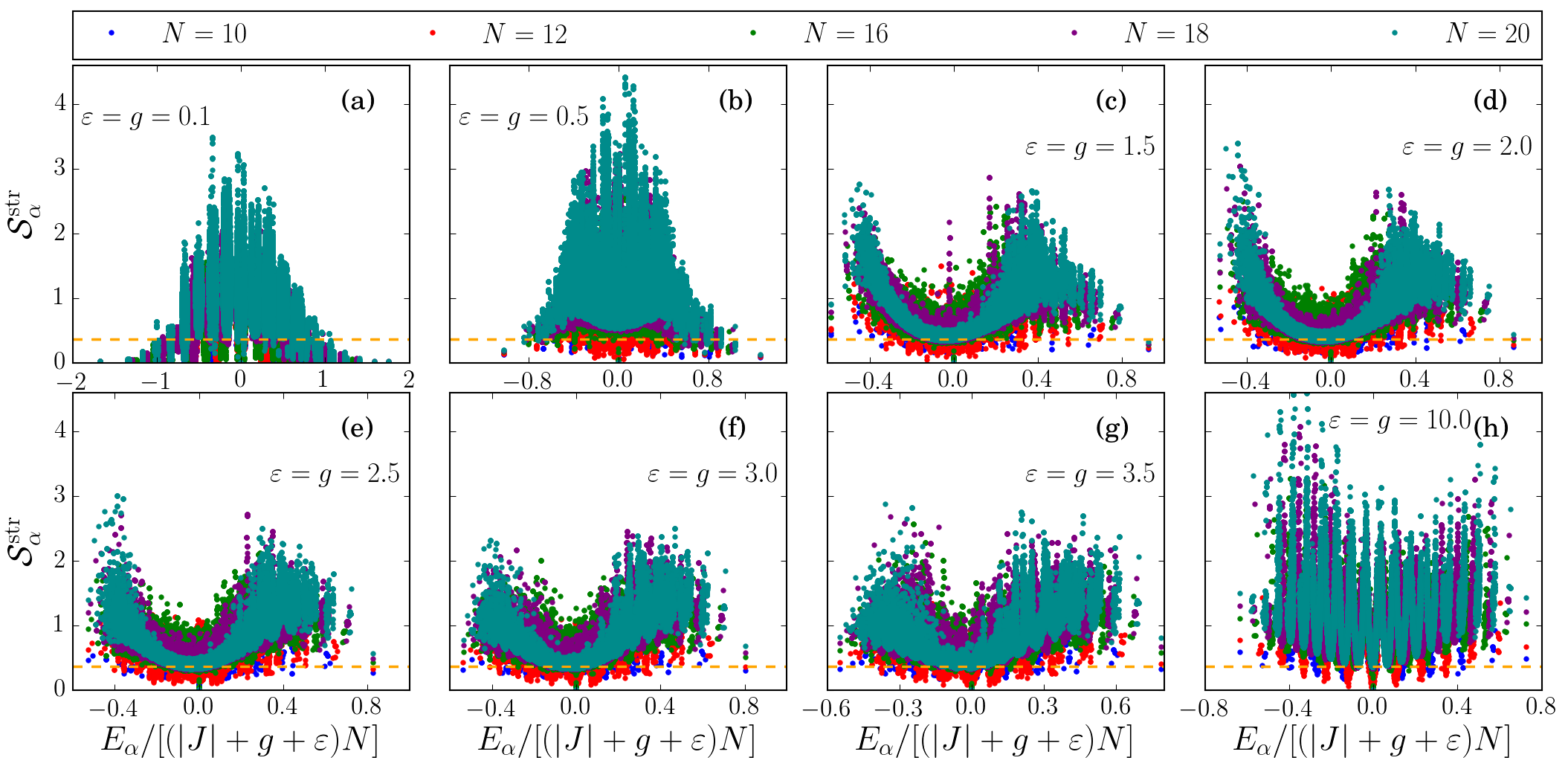}
 \vspace{-0.1cm}
 \caption{(Color online) Same as Fig.~\ref{fig:struct_entropy_F} for the antiferromagnetic 2D-TFIM ($J=1$) with a longitudinal field of strength $\varepsilon=g$.}
  \label{fig:struct_entropy_AF}
\end{figure*}

\subsection{Delocalization of eigenvectors}

An understanding of how quantum chaos onsets in different parts of the spectrum can be gained by studying the delocalization of the energy eigenstates in the basis used to diagonalize the Hamiltonian \cite{santos_rigol_10a,santos_rigol_10b}. Let \{$|\alpha\rangle$\} be the eigenstates of the Hamiltonian in a given symmetry sector, and \{$|m\rangle$\} be the computational basis used in that sector, i.e., $|\alpha\rangle=\sum_m c_m^\alpha|m\rangle$, where the sum runs over the ${\cal D}$ states that make that particular symmetry sector. The amount of delocalization in the computational basis is usually measured using two quantities, the Shannon (information) entropy 
\begin{equation}
 {\cal S}_\alpha^\text{inf} \equiv -\sum_m |c_m^\alpha|^2\ln(|c_m^\alpha|^2),
\end{equation}
and the inverse participation ratio (IPR), 
\begin{equation}
 \xi_\alpha \equiv \frac{1}{\sum_m |c_m^\alpha|^4}.
\end{equation}
Within the GOE, these delocalization indicators are predicted to be: $S^\text{inf}_{\rm{GOE}} \simeq \ln(0.48{\cal D})$ and $\rm{IPR}_{\text{GOE}}\simeq{\cal D}/3$ \cite{izrailev1990simple, zelevinsky1996nuclear}; i.e., they depend on ${\cal D}$. 

Since here we are dealing with symmetry sectors with a wide range of dimensionalities, and for some of them we do not even resolve all space symmetries, a better quantity to characterize the onset of quantum chaos is the structural entropy \cite{santos_rigol_10b}. It is defined as \cite{pipek_varga_92,jacquod_varga_02}
\begin{equation}
{\cal S}^{\text{str}}_\alpha  \equiv {\cal S}_\alpha^\text{inf} - \ln \xi_\alpha.
\label{Sstructural}
\end{equation}
Within the GOE: ${\cal S}^{\text{str}}_\text{GOE} \approx 0.3646$; i.e., it is, to leading order, independent of ${\cal D}$. Hence, this quantity allows one to compare eigenvectors in different symmetry sectors without the need of extra manipulations \cite{santos_rigol_10b}.

\begin{figure}[!tb] 
  \includegraphics[width=0.99\columnwidth]{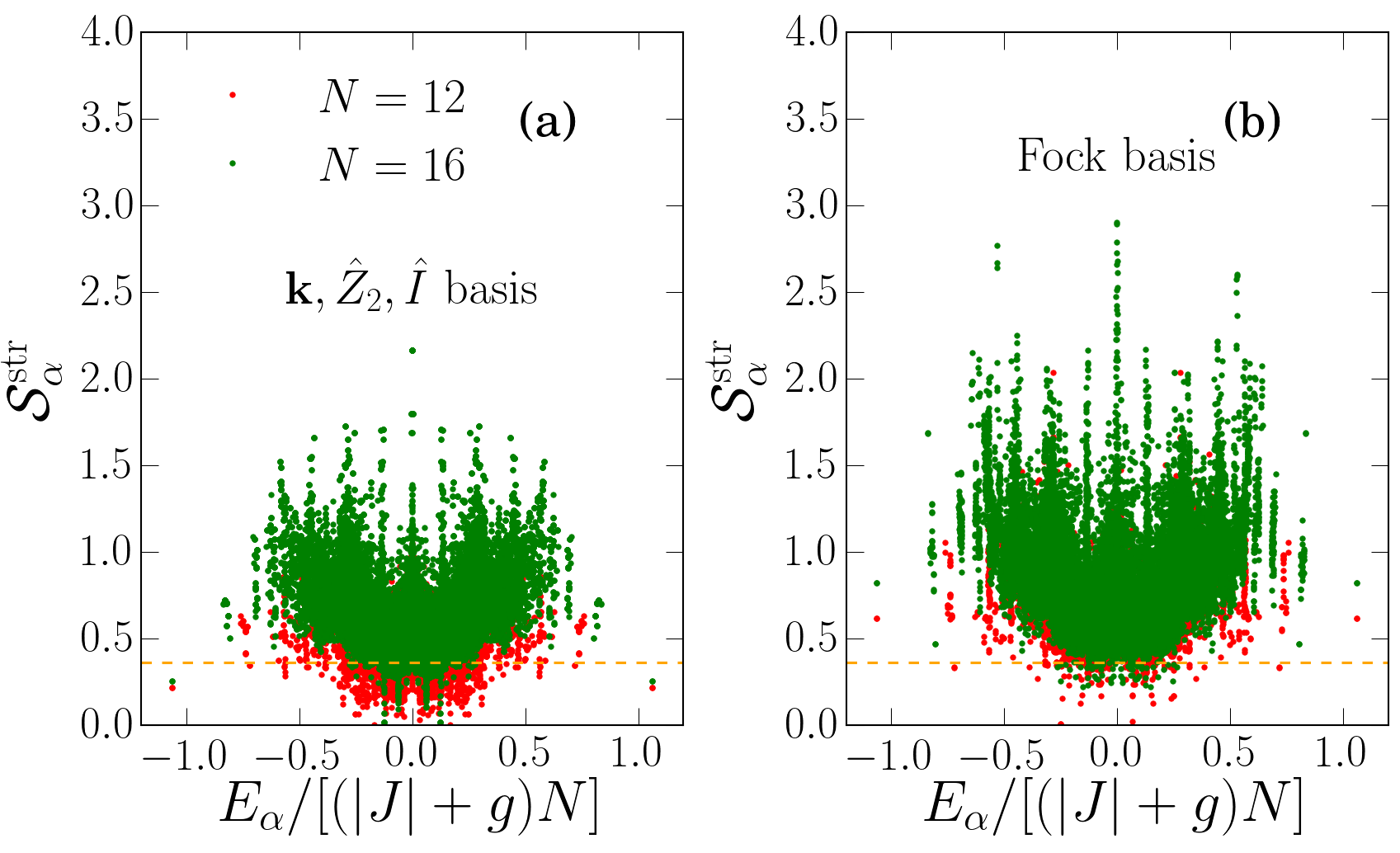}
 \vspace{-0.1cm}
 \caption{(Color online) Structural entropy of the ferromagnetic model with $g=1.0$ ($\varepsilon=0$) for $N=12$ and 16 sites. (a) Results after accounting for translational, $Z_2$, and inversion symmetry (when present).
 (b) No symmetry is used when fully diagonalizing the Hamiltonian.}
  \label{fig:struct_entropy_F_16sites_comparison}
\end{figure}

In Figs.~\ref{fig:struct_entropy_F} and \ref{fig:struct_entropy_AF}, we show the structural entropy for the ferromagnetic and antiferromagnetic models, respectively, for five different systems sizes and eight values of the transverse field. We note that, for each system size, the results obtained for all symmetry sectors (as per Table~\ref{table:dimensions}) are reported using the same symbol. The results in Fig.~\ref{fig:struct_entropy_F} and \ref{fig:struct_entropy_AF} are qualitatively similar. As one departs from the integrable limits $g=0$ and $g=\infty$, and as one increases the system size, the structural entropy away from the edges of the spectrum becomes a smoother function of the energy of the eigenstates. This is a clear signature of quantum chaos. The narrowest support for ${\cal S}^{\text{str}}_\alpha$ within a small energy window in the middle of the spectrum is seen in Fig.~\ref{fig:struct_entropy_AF} when $\varepsilon=g\approx2$. In general, the results for the antiferromagnetic model are slightly better than for the ferromagnetic one. This is understandable as, for any given system size, the former has less symmetries.

Our results support the conclusion in Ref.~\cite{santos_rigol_10b} that the structural entropy is a useful quantity to detect quantum chaos in systems with unaccounted symmetries. To make this point even clearer, in Fig.~\ref{fig:struct_entropy_F_16sites_comparison} we compare the structural entropy of the ferromagnetic 2D-TFIM ($g=1$) for systems with $N=12$ and 16 sites when: (a) one accounts for translational, $Z_2$, and inversion symmetry (when present), and (b) one does not resolve any symmetry (in which case we can fully diagonalize the Hamiltonian only up to $N=16$). While numerical degeneracies lead to obvious quantitative differences between panels (a) and (b), the results are qualitatively similar and, with increasing system size, one could potentially identify that there is quantum chaos in the system even if one does not resolve any of the symmetries of the model.  

\section{Eigenstate expectation values}\label{sec:sec4}

\begin{figure*}[!t] 
 \includegraphics[width=1.9\columnwidth]{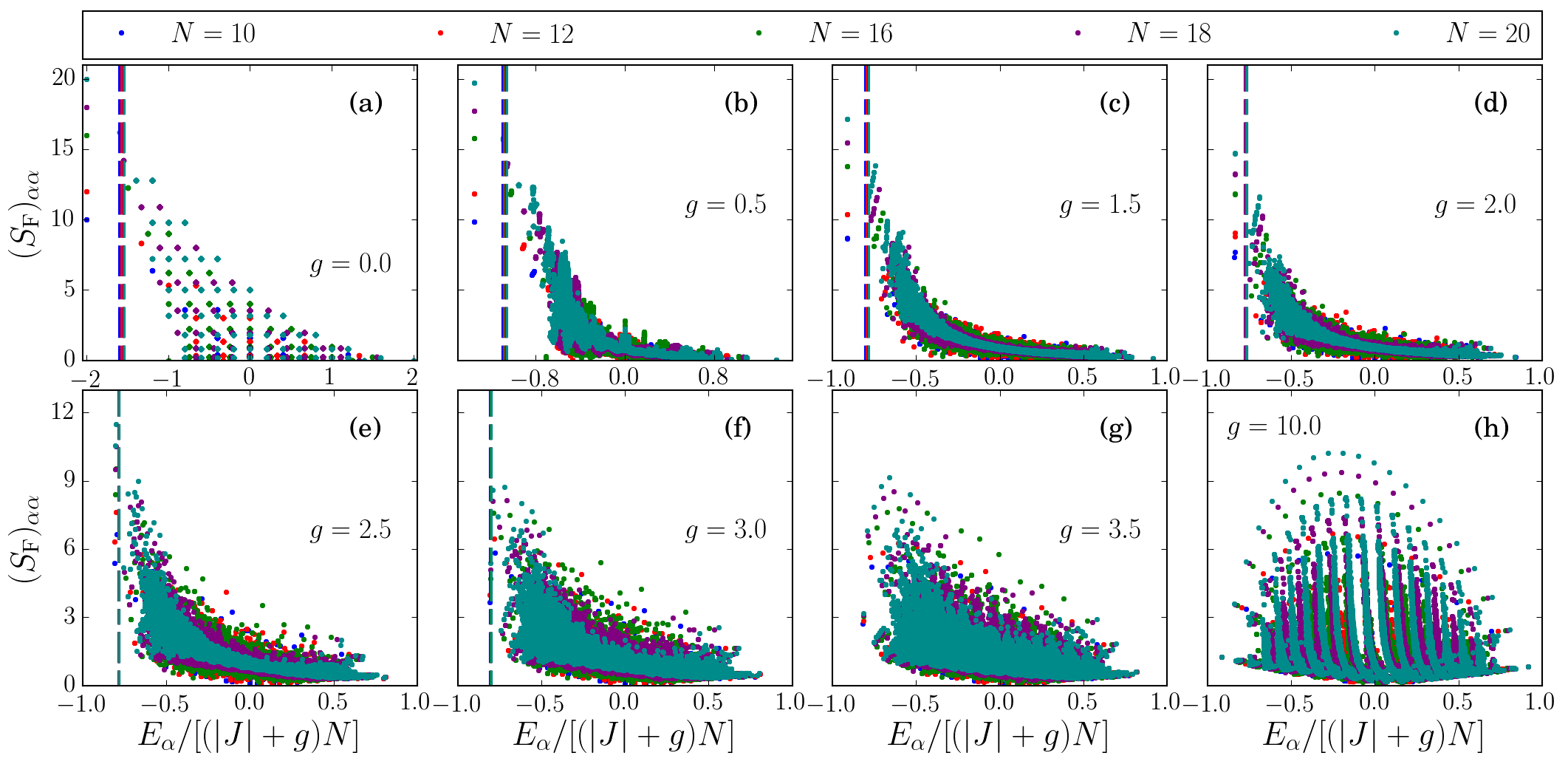}
 \vspace{-0.1cm}
 \caption{(Color online) Energy-eigenstate expectation values of the ferromagnetic structure factor, $(S_\text{F})_{\alpha\alpha}= \langle\alpha|\hat S_\text{F}|\alpha\rangle$, in the ferromagnetic 2D-TFIM ($\varepsilon=0$). The narrowing of the support of the eigenstate expectation values with increasing system size is an indication of the occurrence of eigenstate thermalization. Vertical dashed lines depict the critical energies $E_c$ (Eq.~\ref{eq:energy_temp}) below which the system is expected to display long range order.}
  \label{fig:S_F}
\end{figure*}
\begin{figure*}[!t] 
 \includegraphics[width=1.9\columnwidth]{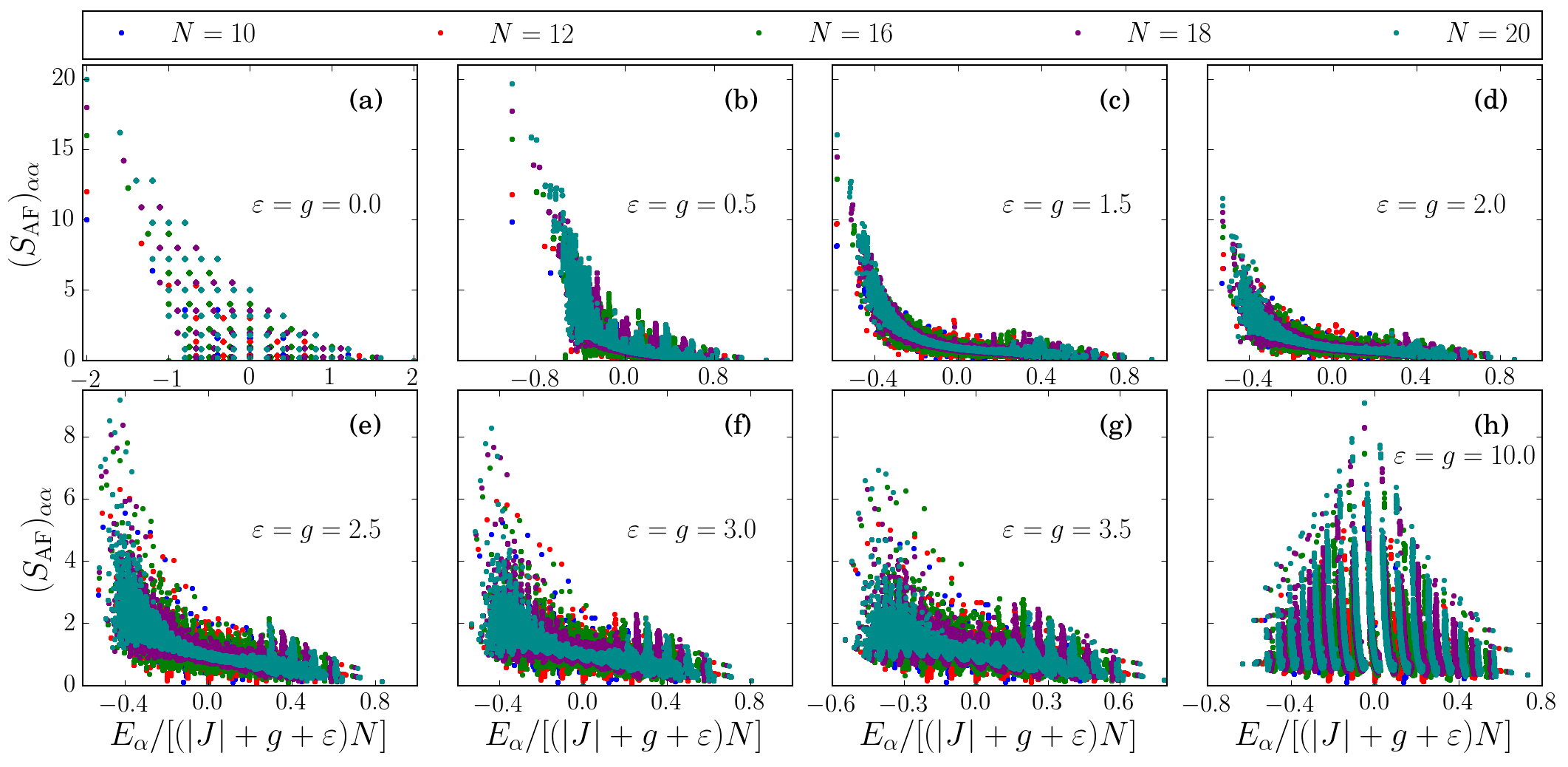}
 \vspace{-0.1cm}
 \caption{(Color online) Energy-eigenstate expectation values of the antiferromagnetic structure factor, $(S_\text{AF})_{\alpha\alpha}= \langle\alpha|\hat S_\text{AF}|\alpha\rangle$, in the antiferromagnetic 2D-TFIM with a longitudinal field of strength $\varepsilon=g$. As for the ferromagnetic case, the narrowing of the support of the eigenstate expectation values with increasing system size is an indication of the occurrence of eigenstate thermalization.}
  \label{fig:S_AF}
\end{figure*}

In order to check whether eigenstate thermalization occurs in the models studied in Sec.~\ref{sec:sec4}, we compute the energy-eigenstate expectation values of two operators that can be used to detect long-range order in those models. For the ferromagnetic one, we compute the energy-eigenstate expectation values of the ferromagnetic structure factor
\begin{equation}
 \hat S_\text{F} = \frac{1}{N}\sum_{{\bf i},{\bf j}}{\hat \sigma_{\bf i}^z}{\hat \sigma_{\bf j}^z}.
\end{equation}
Analogously, for the antiferromagnetic model, we compute the energy-eigenstate expectation values of the antiferromagnetic structure factor
\begin{equation}
 \hat S_\text{AF} = \frac{1}{N}\sum_{{\bf i},{\bf j}}(-1)^{\theta_{{\bf ij}}}\,{\hat \sigma_{\bf i}^z}{\hat \sigma_{\bf j}^z},
\end{equation}
where $\theta_{{\bf ij}}=1$ if ${\bf i}$ and ${\bf j}$ belong to the same sublattice of the bipartite square lattice, and $\theta_{{\bf ij}}=-1$ otherwise. (Note that these two quantities are invariant under the $Z_2$ symmetry operation mentioned before.) In the ordered phase, these two quantities are proportional to $N$, while in the paramagnetic phase they are ${\mathcal O}(1)$.

Figures~\ref{fig:S_F} and \ref{fig:S_AF} show the eigenstate expectation values of the ferromagnetic and antiferromagnetic structure factors in the ferromagnetic and antiferromagnetic 2D-TFIMs, respectively, as computed in all the eigenstates of the Hamiltonian. As for the structural entropy in Figs.~\ref{fig:struct_entropy_F} and \ref{fig:struct_entropy_AF}, one can see that as one departs from the integrable limits $g=0$ and $g=\infty$, and as one increases the system size, the eigenstate expectation values away from the edges of the spectrum become a smoother function of the eigenstate energies. This is a clear indication of the occurrence of eigenstate thermalization. Similarly to the results for the structural entropy (though maybe slightly less obvious), the narrowest supports for the eigenstate expectation values are obtained for the antiferromagnetic case, which has the least symmetries.

Next, we attempt to address whether the eigenstate expectation values of the structure factors in the ordered phases exhibit eigenstate thermalization. In order to do that, we need to identify which eigenstates fall in the part of the spectrum that exhibits long-range order. This can be done using the critical temperature for the phase transition $T_c$. Given $T_c$, one can calculate the mean energy of the system, $E_c$, at that temperature:
\begin{equation}
 E_c=\frac{\sum_\alpha E_\alpha \exp(-E_\alpha/T_c)}{\sum_\alpha \exp(-E_\alpha/T_c)},
 \label{eq:energy_temp}
\end{equation}
where we have set the Boltzmann constant to one. One can then say that, as the system size increases, the eigenstates with energies $E_\alpha<E_c$ fall in the part of the spectrum that exhibits long-range order. 

The ferromagnetic 2D-TFIM has been intensively studied in the past (Refs.~\cite{elliot_pfeuty_wood_70, elliot_wood_71, pfeuty_elliot_71, leeuwen1998dmrg}). Its finite temperature phase diagram was computed in a pioneering series expansions study \cite{elliot_wood_71}, and has been corroborated using quantum Monte Carlo simulations \cite{nagai1987,suzuki2012}. Using the results for $T_c(g)$ from the latter study, we have calculate $E_c(g)$ in all clusters (for $g<3.044$, which is the critical value for the ground-state phase transition). The results obtained for $E_c(g)$ are presented in Fig.~\ref{fig:S_F} as dashed lines. These estimates are significantly lower than those made in Ref.~\cite{FratusSrednicki} using fluctuation-corrected mean-field theory \cite{Stratt1986}, indicating that much of the branch structure for the magnetization seen in Ref.~\cite{FratusSrednicki} actually occurs in the disordered phase. We find that, for the system sizes accessible to us via full exact diagonalization, only a few states reside in the ordered phase. Therefore it is not possible for us to make a definitive statement about the appearance of eigenstate thermalization in the ordered phase of the spectrum.

We are not aware of studies of the phase diagram of the antiferromagnetic 2D-TFIM with a longitudinal field $\varepsilon=g$. Because of this, its $T_c(g)$ is not known to us, and we are not able to report results for $E_c(g)$ as we do for the ferromagnetic case.

\subsection{Scaling with system size}

Next we address how the eigenstate to eigenstate fluctuations in the expectation value of the structure factors scale with increasing system size. We compute 
\begin{equation}\label{eq:differr}
 (\Delta S_\text{F})_\alpha\equiv|(S_\text{F})_{\alpha+1,\alpha+1}-(S_\text{F})_{\alpha,\alpha}|
\end{equation}
for the ferromagnetic 2D-TFIM and 
\begin{equation}\label{eq:difaferr}
(\Delta S_\text{AF})_\alpha\equiv|(S_\text{AF})_{\alpha+1,\alpha+1}-(S_\text{AF})_{\alpha,\alpha}| 
\end{equation}
for the antiferromagnetic 2D-TFIM with a longitudinal field. We stress that to compute these quantities we order {\it all} the energy eigenstates with increasing energy. For that, we collect the results from all sectors that are diagonalized independently, i.e., the entire spectrum is put together into a single ordered list before calculating Eqs.~\eqref{eq:differr} and \eqref{eq:difaferr}. From the eigenstate thermalization hypothesis (ETH) \cite{mrigolreview2015}, one expects the maximal values of $(\Delta S_\text{F})_\alpha$ and $(\Delta S_\text{AF})_\alpha$ to decrease exponentially with system size. In Ref.~\cite{kim_14}, this was shown to be the case for observables in various one-dimensional models (including the TFIM with a longitudinal field) when taking the central half of the energy eigenstates. As the system size increases, this is a statement about eigenstates whose energies are that of a thermal ensemble at infinite temperature, which constitute the overwhelming majority of states in the spectrum of large systems.

\begin{figure}[!t] 
 \includegraphics[width=0.99\columnwidth]{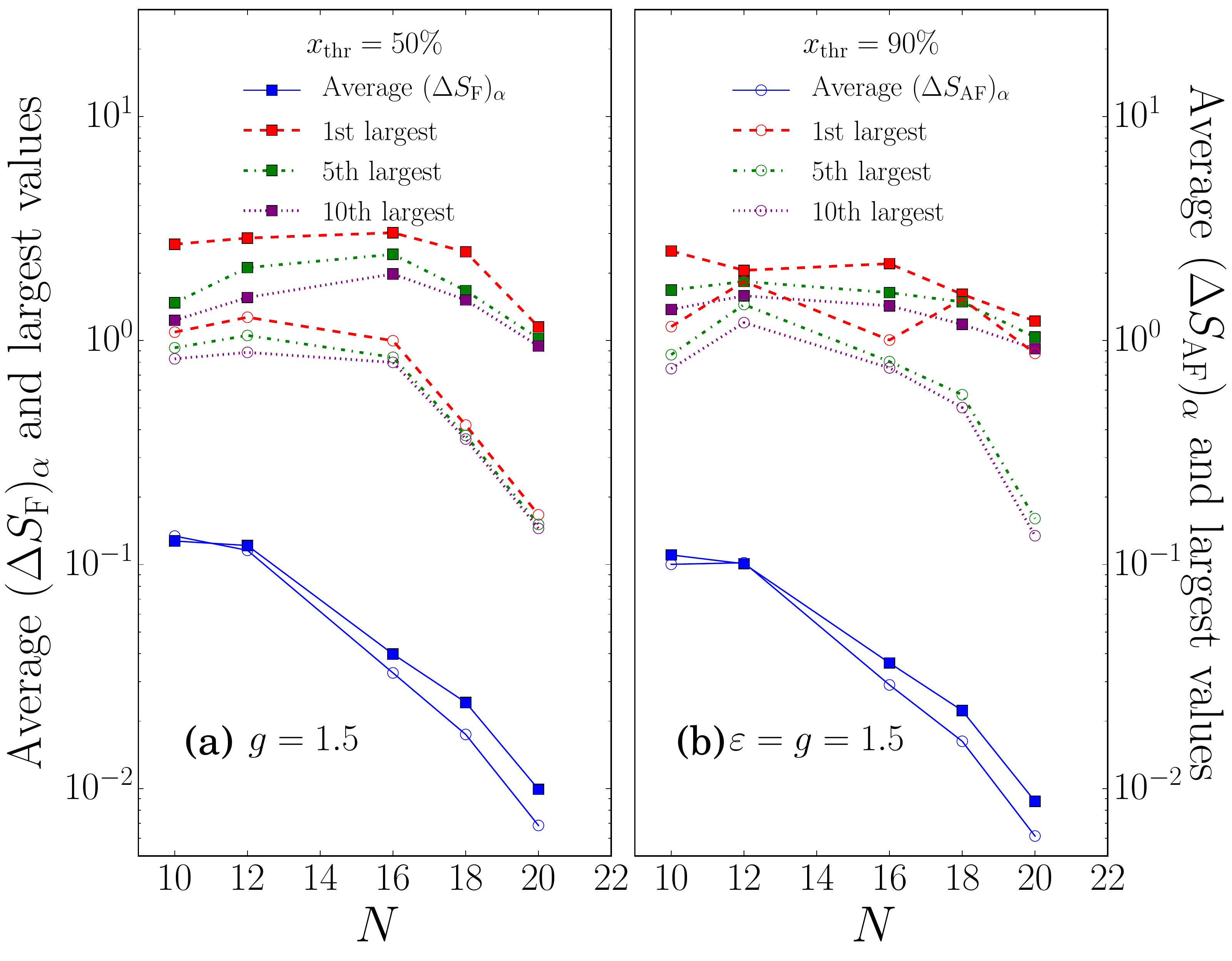}
 \vspace{-0.1cm}
 \caption{(Color online) Largest, fifth largest, tenth largest, and average value of: (a) $(\Delta S_\text{F})_\alpha$ for the ferromagnetic 2D-TFIM ($g=1.5$ and $\varepsilon=0$) and (b) $(\Delta S_\text{AF})_\alpha$ for the antiferromagnetic 2D-TFIM in the presence of a longitudinal field with $\varepsilon=g=1.5$, plotted as a function of the number of lattice sites in the cluster. All those quantities are computed within two windows of eigenstates characterized by $x_\text{thr}$ (see text). All those quantities are computed
within two windows of eigenstates characterized by $x_\text{thr} = 50\%$  (filled symbols) and $x_\text{thr} = 90\%$ (open symbols). See text for the definition of $x_\text{thr}$.}
  \label{fig:diff_EEV_AF_F}
\end{figure}

In order to make a stronger statement about the eigenstate to eigenstate fluctuations, we compute their largest values, as well as their average, after removing all states with energy $E_\alpha$ such that $(E_\alpha-E_0)/|E_0|<x_\text{thr}$ ($E_0$ is the ground state energy) and $(E_{\mathcal D}-E_{\alpha})/E_{\mathcal D}<x_\text{thr}$ ($E_{\mathcal D}$ is the eigenstate with the highest energy in the spectrum). States at the edges of the spectrum need to be removed because, as mentioned before, they neither exhibit quantum chaos nor eigenstate thermalization. So as long as $x_\text{thr}\ncong 1$, our statements about the eigenstate to eigenstate fluctuations are not restricted to eigenstates whose energy is that of a thermal ensemble at infinite temperature (for which $E_{\alpha}\cong0$ and $x_\text{thr}\cong 1$).

In Figs.~\ref{fig:diff_EEV_AF_F}(a) and \ref{fig:diff_EEV_AF_F}(b) we plot the results obtained for $(\Delta S_\text{F})_\alpha$ and $(\Delta S_\text{AF})_\alpha$, respectively, as a function of the number of lattice sites for two values of $x_\text{thr}$. We report results for the largest, the fifth largest, and the tenth largest values of those quantities in the windows selected, as well as the average value (which is dominated by the aforementioned ``infinite-temperature'' states). The decrease of the average value is consistent with an exponential for the systems with $N\geq 12$, independent of the value of $x_\text{thr}$. For the extremal values, on the other hand, the onset of the exponential decrease requires larger lattices and is better seen for $x_\text{thr}=0.9$. 

\begin{figure}[!t] 
 \includegraphics[width=0.99\columnwidth]{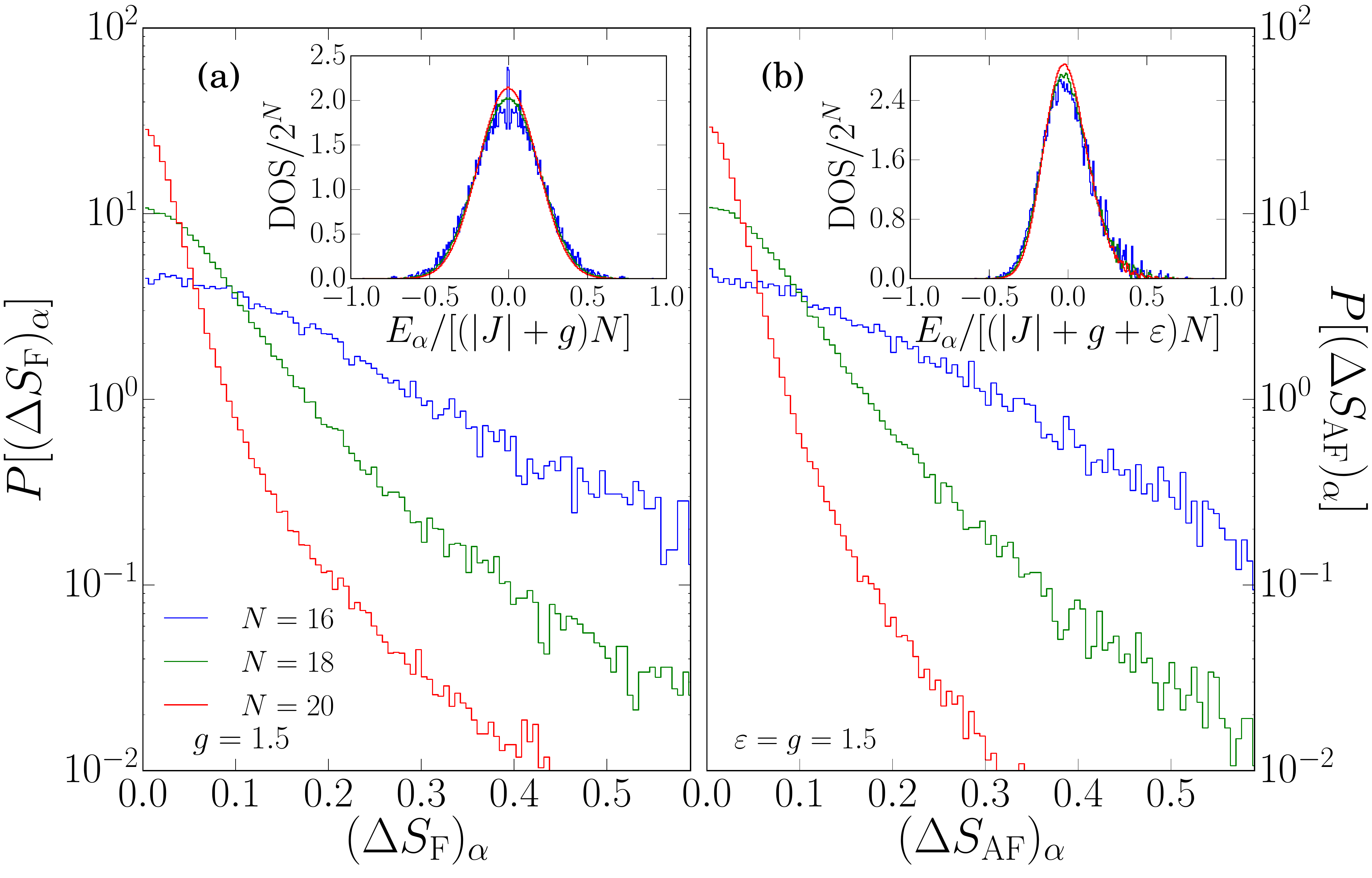}
 \vspace{-0.1cm}
 \caption{(Color online) Distribution of: (a) $(\Delta S_\text{F})_\alpha$ for the ferromagnetic 2D-TFIM ($g=1.5$ and $\varepsilon=0$) and (b) $(\Delta S_\text{AF})_\alpha$ for the antiferromagnetic 2D-TFIM in the presence of a longitudinal field with $\varepsilon=g=1.5$. The distributions were computed for $x_\text{thr}=0.5$. (Insets) Density of states in the clusters.}
  \label{fig:distribution_consecutive_EEVs}
\end{figure}

The distribution of values of $(\Delta S_\text{F})_\alpha$ and $(\Delta S_\text{AF})_\alpha$, for $x_\text{thr}=0.5$, is shown in Figs.~\ref{fig:distribution_consecutive_EEVs}(a) and \ref{fig:distribution_consecutive_EEVs}(b), respectively. The results for both quantities are not only qualitatively but also quantitatively similar. One can see that, as expected, the distributions become increasingly peaked about $(\Delta S_\text{F})_\alpha=(\Delta S_\text{AF})_\alpha=0$ as the system size increases, and their support decreases significantly (consistent with decreasing exponentially fast) as the system size is increases. The exponential increase of the density of states with increasing system size, as well as the Gaussian nature of the density of states in the systems studied here, can be seen in the insets in Fig.~\ref{fig:distribution_consecutive_EEVs}.

\section{Summary}\label{sec:sec5}

We have systematically studied quantum chaos indicators and energy-eigenstate expectation values of structure factors in the ferromagnetic 2D-TFIM, and the antiferromagnetic 2D-TFIM in the presence of a longitudinal field, in the square lattice. We have shown how quantum chaos and eigenstate thermalization onset in those systems as one departs from integrable limits and increases the system size. While many systematic studies of these topics have been undertaken in one-dimensional lattices \cite{rigol_09a, rigol_09b, santos_rigol_10a, santos_rigol_10b, neuenhahn_marquardt_12, genway_ho_12, khatami_pupillo_13, beugeling_moessner_14, kim_14, sorg_vidmar_14}, this is among the first to be carried out in two dimensions, for which scaling analyses are very challenging due to the fast increase of the Hilbert space with the linear dimension of the system. We leave open the questions of whether quantum chaos and eigenstate thermalization occur in eigenstates of a Hamiltonian that exhibit long range order. Answering those questions appears challenging to full exact diagonalization studies and other computational techniques might be needed to address them.

\begin{acknowledgments}
This work was supported by CNPq (R.M.), NSF Grant PHY13-16748 (K.F. and M.S.), and the Office of Naval Research (M.R.). The computations were performed in the Institute for CyberScience at Penn State, the Center for High-Performance Computing at the University of Southern California, and CENAPAD-SP.\\
\end{acknowledgments}


\section{Appendix}

\begin{figure}[!th] 
 \includegraphics[width=0.99\columnwidth]{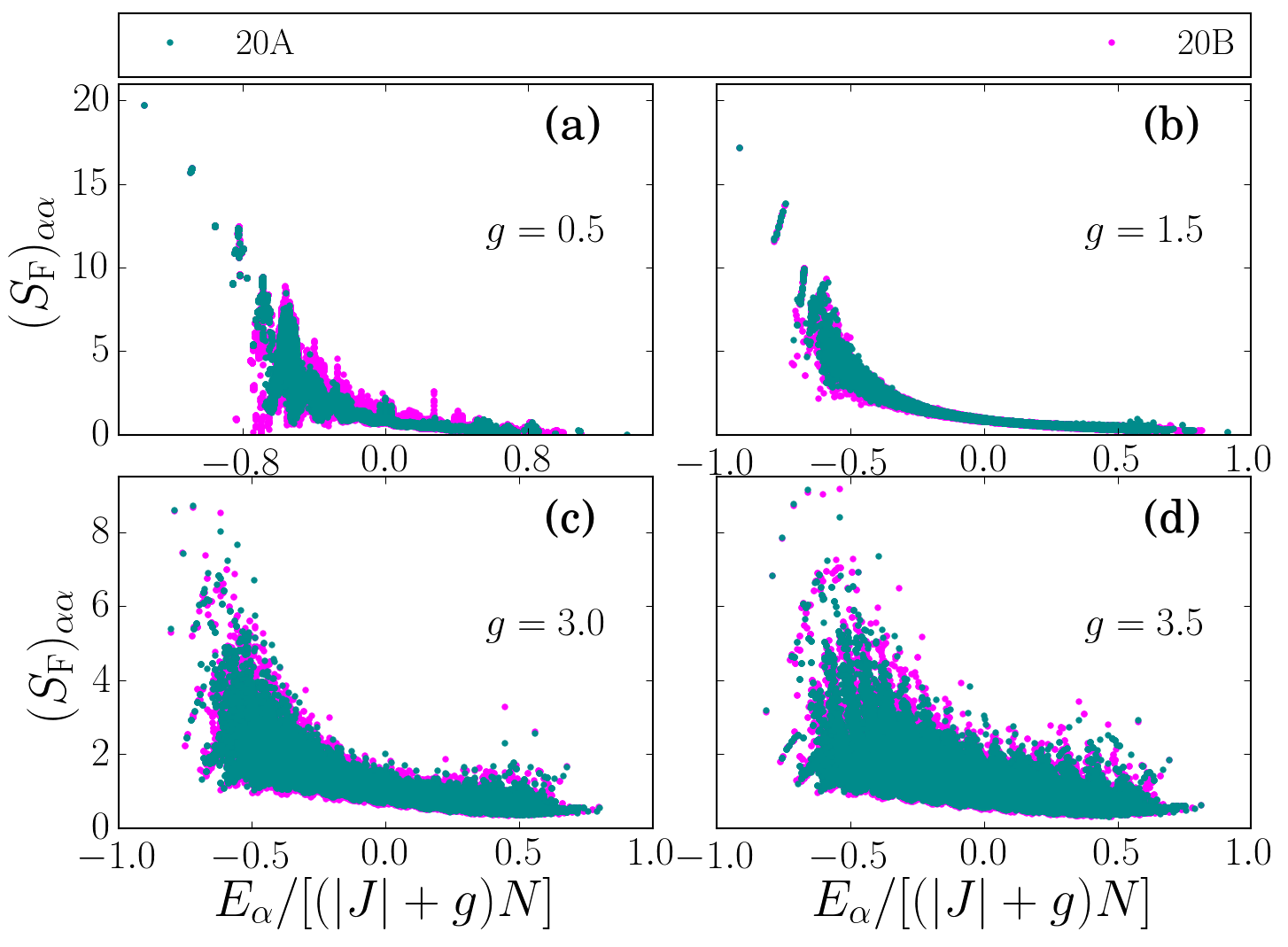}
 \vspace{-0.1cm}
 \caption{(Color online) Energy-eigenstate expectation values of the ferromagnetic structure factor, $(S_\text{F})_{\alpha\alpha}= \langle\alpha|\hat S_\text{F}|\alpha\rangle$, in the ferromagnetic 2D-TFIM ($\varepsilon=0$) for the two clusters with $N=20$ (20A and 20B, see Fig.~\ref{fig:tilted_square}). For all the values of the transverse field depicted, the support of the eigenstate expectation values is narrower in cluster 20A. One can then conclude that this cluster suffers from smaller finite size effects than cluster 20B. All results shown in the main text for $N=20$ are for cluster 20A.}
  \label{fig:S_F_20A_20B_comparison}
\end{figure}

The narrowing of the support of the energy-eigenstate expectation values of few-body operators with increasing system size is a direct consequence of the occurrence of eigenstate thermalization. We should stress, however, that clusters with the same number of sites but different geometries can display differences in the energy-eigenstate expectation values. Figure \ref{fig:S_F_20A_20B_comparison} shows the energy-eigenstate expectation values of $\hat S_\text{F}$ for the clusters 20A and 20B (see Fig.~\ref{fig:tilted_square}) within the ferromagnetic 2D-TFIM. One can see that the eigenstate to eigenstate fluctuations of the expectation values is larger in cluster 20B than in 20A, i.e., the former suffers from stronger finite size effects. Because of this, in the main text we showed results only for cluster 20A.

\bibliography{transverse_ising}

\end{document}